%% file: main.tex
\documentclass[10pt, conference]{IEEEtran}
\input{writing-support}

\begin{document}
\title{\depminer: A Pipelineable Tool for Mining of Intra-Project Dependencies 
}

\author{\IEEEauthorblockN{Vladislav Repinskiy}
\IEEEauthorblockA{\textit{Department of Computer Science} \\
\textit{University College London}\\
London, England \\
vlad.repinskiy.17@ucl.ac.uk}
\and
\IEEEauthorblockN{Vladimir Kovalenko}
\IEEEauthorblockA{\textit{Intelligent Collaboration Tools Lab} \\
\textit{JetBrains Research}\\
Amsterdam, The Netherlands \\
vladimir.kovalenko@jetbrains.com}
}

\maketitle

\begin{abstract}
Dependency analysis is recognised as an important field of software engineering due to a variety of reasons. There exists 
% of appropriate dependency management in practice is substantiated by 
a large pool of tools providing assistance to software developers and architects. Analysis of inter- and intra-project dependencies can help provide various insights about the entire development process. 
% but despite that fact, these days theoretical dependency analysis is a rare sight. 
There is currently a lack of tools that would support researchers by extracting intra-project dependencies data in a format most suited for further analysis. 

In this paper we introduce \depminer\footnote{Anonymous. (2021, January 13). DepMiner. Zenodo. http://doi.org/10.5281/zenodo.4441367}~--- an open source, language-agnostic tool for mining detailed dependencies data from source code, based on extensive static analysis capabilities of an industry standard IDE. \depminer can be easily integrated into arbitrary mining pipelines to conduct large-scale source code processing jobs involving intra-project dependencies. It is easily extensible to support other languages of source code, different granularities of analysis, and other use-specific needs.
\end{abstract}

\section{Introduction}
\input{sections/intro}

\section{Background and Motivation}
\input{sections/background}

\section{IntelliJ Platform}
\input{sections/platform_overview}

\section{DepMiner Design}
\input{sections/depminer_design}

\section{Contribution, quality, and distribution}
\input{sections/evaluation}

\bibliographystyle{IEEEtran}
\bibliography{bibliography}
\end{document}

%% file: writing-support.tex
\usepackage{cite}
\usepackage{amsmath,amssymb,amsfonts}
\usepackage{algorithmic}
\usepackage{graphicx}
\usepackage{textcomp}
\usepackage{xcolor}
\usepackage{soul}
\usepackage{url}
\usepackage{xspace}
\usepackage{hyperref}

\def\BibTeX{{\rm B\kern-.05em{\sc i\kern-.025em b}\kern-.08em
    T\kern-.1667em\lower.7ex\hbox{E}\kern-.125emX}}
    
\newcommand{\todo}[1] {}
\newcommand{\depminer}{DepMiner\xspace}

%% file: sections/intro.tex
Dependency analysis is a major topic in software engineering.
The importance of dependency analysis comes from the variety of its applications in software maintenance and quality control. 
Dependencies data can be extracted and analysed to help ensure high standards of code quality, performance, and maintainability. 
Information about dependencies is used in mature code analysis tools~\cite{cppdepend, jdepend} as well as in the development of new software engineering techniques in research settings~\cite{Zimmermann2008, Nguyen2010, Csaba2013, Dietrich2008, Mohammadi2019}.

Extracting dependencies information from source code is challenging, as it involves building comprehensive code analysis pipelines.
Technical work required to build such pipelines builds up to a substantial part of work in research settings.
With no adaptive, reusable toolkit openly available for extraction of dependencies data, technical work of building mining pipelines diverts the focus of researchers from problems in their primary domain.
However, reliable infrastructure for polyglot source code analysis exists: it is one of the pillars of modern IDEs. 
Reusing language analysis capabilities of IDEs for data collection tasks looks like the missing link towards drastically reducing the technical complexity of data mining in many areas of software engineering research.

In this paper we present \depminer~-- an open source, reusable, extensible mining tool that uses static analysis capabilities of the IntelliJ platform~\cite{intellijplatform} to provide an 
% adaptive and extensible (было выше)
instrument for mining intra-project dependencies data from source code. 
\depminer is designed to cater for a wide variety of mining tasks and is easy to integrate into existing pipelines. 
In addition, we discuss the core features of the IntelliJ Platform relevant for data collection, and then describe the internals of \depminer.
We include a demonstration of \depminer's use in the form of a web application. \footnote{\depminer Example Use Case. Anonymous. (2021, January 15). DepMiner Visualization App. Zenodo. http://doi.org/10.5281/zenodo.4441366}

%% file: sections/background.tex
\subsection{Dependency Analysis}
Ideally, the system architecture is thoroughly developed according to the project goals and requirements before the actual development phase takes place~\cite{Garlan1995, Perry1992}. 
This approach allows more precise project management in the future stages and helps achieve higher standards of clarity in communication between all stakeholders involved~\cite{Pich2008}. 

Despite the universal acceptance of these facts in industry, initial design decisions might be overturned with time, due to the unforeseen problems and limitations or due to poor development practice. 
The erosion of clean structure of a codebase is common~\cite{Zapalowski2018}, and in some cases it might weaken and even destroy the initial intent completely~\cite{Garlan1995}.
Furthermore, unintended design alterations induce the accumulation of technical debt, which may make the current development and maintenance processes more convoluted, if left unchecked. 

Dependency analysis provides a way to monitor, address and contain such problems on all stages of system development~\cite{Nord2012, Zhang2006, Leitch2003}. 
Development of the visual representation of software architecture~\cite{Brondum2012} such as dependency graphs~\cite{Pinzger2008} are a common motive for engineering and scientific investigations.

Besides use in software maintenance, dependency graphs are a vital source of data in development of novel software engineering techniques in research settings.
Prior work demonstrates that the network theory analysis of dependency graphs can help predict software defects~\cite{Zimmermann2008} and vulnerable code components~\cite{Nguyen2010}.
Another popular application of dependency graphs data is cluster analysis and the clustering of software in general.
It is believed that high clustering plays a key role in modular applications design and architecture, and some studies suggest that clustering is correlated directly to software quality~\cite{Csaba2013}.
Intra-project artefact dependency graphs have been proved crucial in detecting clusters in software \cite{Dietrich2008}, as well as in developing novel automated software clustering algorithms and tools \cite{Mohammadi2019}.
Understanding dependencies data helps researchers throughout the whole variety of applied and pure studies. 

\subsection{Existing Approaches To Dependency Analysis}
Many existing solutions and tools focus on data visualisation, generating UML diagrams and dependency graphs of the projects. Tools like Structure101~\cite{structure101} and MagicDraw~\cite{magicdraw} aid software architects in planning, managing and assessing the architecture, that do not employ actual code analysis.

Others provide object oriented metrics and other quantifiable insights. 
For instance, metrics such as a number of methods and attributed, SLOC and the depth of inheritance tree of Java classes are typically used to address the matter of quality and the conformity of the project to the OOP design principles. 
There are also more general metrics such as cyclomatic complexity of a program. 
A consistent relationship between the usage of object oriented metrics and quality factors has been demonstrated, as observed by Darcy and Kemerer~\cite{Darcy2005}.

Another large portion of these tools focus primarily on visualisation of modular structure of the application, such as Sourcetrail~\cite{sourcetrail} and discontinued OptimalJ~\cite{wiki:optimalj}. 

Numerous instruments checking vulnerabilities in inter-project dependencies provide secure usage of external libraries and packages, for languages most dependent on it. 
Examples of those include OSSIndex~\cite{ossindex}, Dependency-check~\cite{dep-check} and RetireJS~\cite{retirejs}. 
 
The majority of automated tools are language-dependent, meaning that the tool can only function properly on the codebase maintained in a single language, or a collection of similarly constructed languages. 
For instance, JDepend~\cite{jdepend}, JavaNCSS~\cite{javancss} and Dependency Finder~\cite{depfinder} support JVM language codebases, while CppDepend~\cite{cppdepend} works with C/C++ code.

Most modern IDEs also include similar functionality in their toolkit. For instance, IntelliJ IDEA provides reports on clustering and interdependencies of modules within one project, also suggesting and facilitating possible refactoring to restore application modularity without disrupting the functionality. 

\subsection{Motivation}
Despite the widespread adoption of the aforementioned and other complementary instruments in the industry, they fall out of scope for quantitative, low-granularity alternative dependency analysis problems in research since they do now provide clear access to raw dependencies data for potential analysis. 
The majority of works revolving around the topic of inter and intra-project code dependencies attempt to ease the workload of software architects, lead developers and other professionals dealing with software architecture in its full complexity.

% However, the project that preceded the development of DepMiner did not settle at the extraction of key metrics and insights about intra-project dependencies. 
% Extensive search for a tool that would support collection of \textit{raw dependencies data} (data that would describe exact connection between different code elements) revealed that such solutions are hard to come by. 

% Thus, a tool that would extract information about dependencies from source code was necessary. 
% Additionally, extracting the dependencies data from the source code was not the endpoint of the project - on the contrary, the idea was to repeatedly feed dependencies data further as input for further analysis. 
% None of the preexisting solutions discovered in the process seemed tailored to the quite specific need of the project, and hence the 
\todo{list dependency types, note that call graphs are not the only type}
Existing tools for extraction of dependency graphs from static codebases rely on complex language-specific code analysis technology, and are thus not easily extendable to support multiple programming languages. 

DepMiner is a reusable, open-source tool for extraction of call graphs that can be easily incorporated into pipelines for complex experiments, or extended to support new programming languages or other dependency types.

%% file: sections/platform_overview.tex
DepMiner is based on IntelliJ Platform. It consists of an IntelliJ IDEA plugin and infrastructure to run the processing in headless mode, without spinning up the IDE interface. In this section, we provide a brief overview of IntelliJ Platform from the standpoint of use for code analysis tools like DepMiner.

\subsection{IntelliJ Platform SDK}
IntelliJ Platform is an open source system of abstract components comprising an infrastructure that provides rich language tooling support for developers in the form of a documented SDK~\cite{intellijplatform}. 
% The choice of development framework came naturally, as we wanted to replicate functionality that to some extend already existed within IntelliJ-based IDE namely, dependency analysis capabilities of IntelliJ IDEA. 

The IDE itself can analyse project dependencies on different levels of granularity (module, file, line dependencies). 
However, it does not allow users to export this data outside of the IDE. 
Thus, it is not best suited for data mining settings, especially if the dependency analysis should be performed for multiple projects in one run.
% Components of the IntelliJ Platform can be extended to support a variety of languages and functions, which is why many IDEs and IDE extensions and plugins are made using the IntelliJ Platform.
The core idea behind DepMiner is to utilize the capabilities of the IntelliJ Platform for mining of data that is useful in research contexts and would otherwise require building complex dedicated code analysis tools.
% The platform includes robust open APIs “to build standard IDE functionality, such as a project model and a build system”~\cite{intellijplatform}, which enable various code inspections and provide helper methods for static code analysis, which DepMiner utilises. 

\subsection{Program Structure Interface}
\begin{figure}[h]
    \centering
    \includegraphics[width=0.35\textwidth]{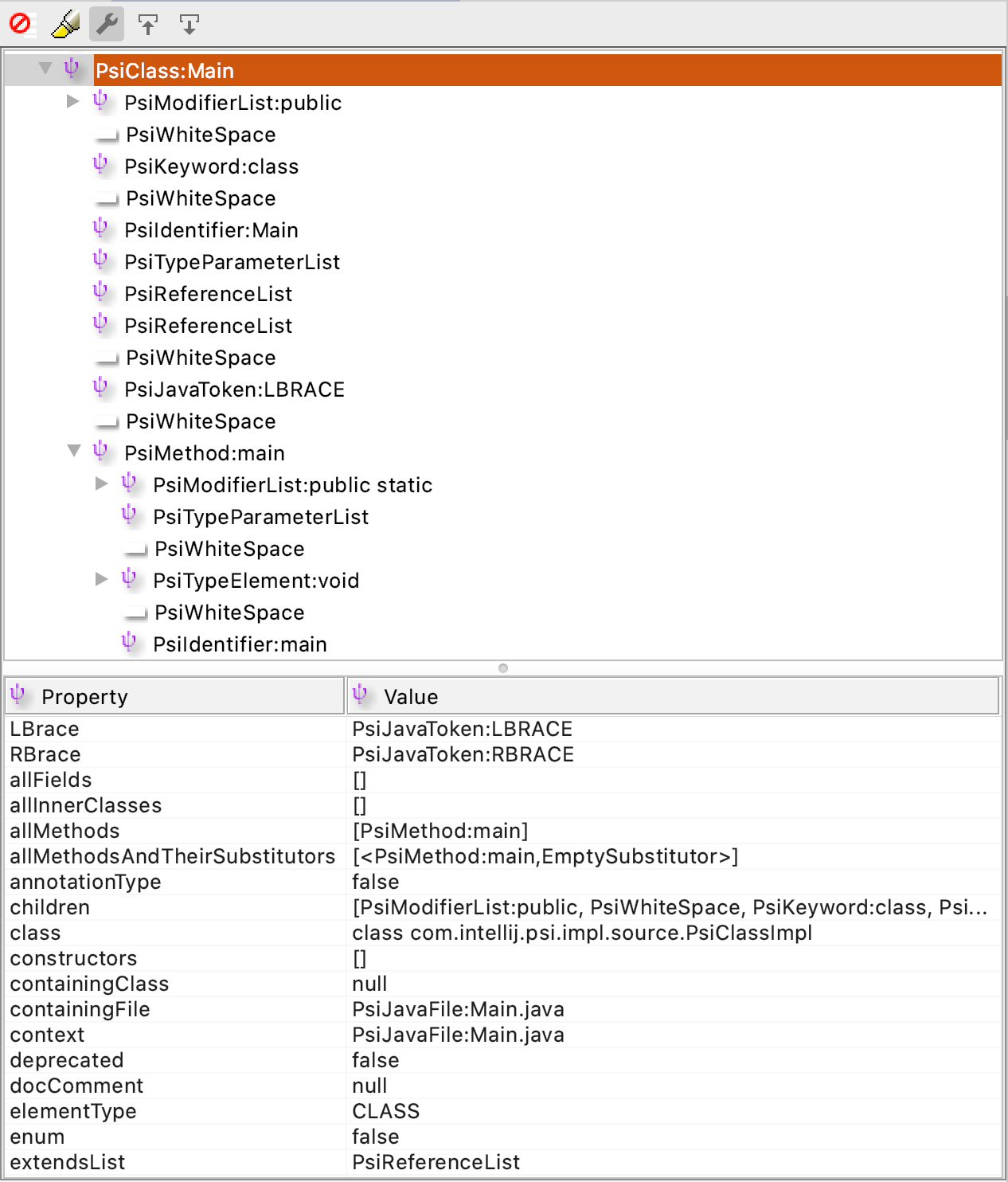}
    \caption{PSI structure of a simple Java file.
    % as seen in the PsiViewer IDEA plugin window. 
    % The tree-like structure of ordering of PSI elements help easy traversal
    }
    \label{fig1}
\end{figure}

One of the pillars of IntelliJ Platform is the Program Structure Interface (PSI).
PSI is a platform layer responsible for representing source code in the form of meaningful semantic models. It is specifically designed to represent the codebase in a way suitable to support IDE code analysis features. 
PSI logic presents any code fragment as a AST-like syntax tree, where each node represents a certain code element and is augmented by numerous fields providing more detail about the element itself. 
There are different types of nodes for different types of code elements. 
For instance, a PSI structure of Java file would normally contain nodes of type PsiClass, PsiMethod, PsiField and so on. 
Figure \ref{fig1} demonstrates a typical PSI structure of a Java file.
Different types of nodes often have different sets of attributes, which in turn provide details of code element’s location in file, containing file name, language and such. 

PSI trees and nodes are constructed on demand, which means that they are only built if and when any component or plugin in the IDE requires the PSI tree of a certain file or code region.
Language-specific PSI parsers are responsible for constructing PSI trees of source code files. IntelliJ IDEA, for instance, can construct PSI trees for JVM-based languages, but it can also be extended with a plugin that would provide support for files in other popular languages. 

The ability to construct PSI trees for source code in different programming languages is implemented as part of language support for IntelliJ-based IDEs as either bundled or optional platform plugins. 
Therefore, relying on PSI, rather than on tailored language-specific parsers, for code analysis helps drastically reduce the complexity of code analysis pipelines.
% IntelliJ Platform provides a method which attempts to construct and return a PSI tree for any valid string in a language recognised by the IDE. 
% This is particularly useful in cases when someone wishes to analyse a file unavailable locally, if the source code can be accessed as a string (for instance, from VCS services).

\subsection{Indexing and References}

Indexing is the framework that lies in depth of dependency resolution in IntelliJ Platform, and thus in DepMiner. 
It provides a mechanism for querying whether a specific code element is present in a file, in a way that is scalable to large codebases performance-wise. 
This same mechanism allows to quickly search for the occurrences of the same token in multiple files.
When a project is opened, the IDE goes through the entire file tree and indexes source code, constructing a map of keys found in each file and their associated values. 
Indices are then serialised and stored in a binary format, to allow efficient access and querying. 
% Indexing helps provide fast response of different features, however it often takes some time to complete on project opening. 
% Before indexes are built, many IDE features are not available, and the IDE is said to be in a “dumb mode”. 
% IntelliJ Platform provides an API to hold certain actions off until project was fully indexed. 

If a PSI element represents a usage of a certain element declared elsewhere in the code, it is linked to the declaration of the element through a PSI reference. However, references are not parts of the PSI tree per se.
PSI elements that represent a usage of another element must provide a PSI reference to the declaration of itself on-demand, using the \texttt{getReferences()} method. 
However, PSI references do not contain actual location of the declaration of the element. 
Instead, they only provide logic on how to locate the declaration of the element for a certain type of element. 
However, in general location resolution is possible under the aegis of the aforementioned mechanism of indices. 

For example, let us consider a PSI element representing a method call in Java. 
On invocation of \texttt{PsiElement.getReferences()}, element logic will retrieve the method name, and query the corresponding index on whether identical tokens exist anywhere in the project. 
Using this data, the platform can figure out whether any other files contain the aforementioned method name and construct corresponding PSI trees. 
Finally, it will traverse the files bottom-up, checking with each element if it is indeed a declaration of the inspected element. 
The process of invoking the logic stored in a certain reference is called resolving, and it can be initiated by calling the \texttt{resolve()} method on a PSI reference.

%% file: sections/depminer_design.tex
\subsection{Requirements}
% This section states the requirements that preceded the development of DepMiner. 
Initially, \depminer was a component of our internal data mining pipeline that we later converted into a reusable tool.
Apart from extraction of dependencies data, the fundamental feature we required of DepMiner is the ability to be integrated into complex data analysis pipelines.
% , which ultimately defined its running interface and output format requirements.

We followed the following requirements 
% in the development process 
to ensure that \depminer can be reused in other settings:

\begin{itemize}
    \item DepMiner must extract dependencies at all levels of granularity from project source code - i.e. tokens, functions and classes encapsulating them, files and directories.
    \item Must run from command line
    \item Must output results in JSON format
    \item Must support projects written in popular languages - Java and Python
    \item Should be extensible to support other languages
    \item Dependencies information must contain information about location and type of code elements
\end{itemize}

\subsection{DepMiner Data Model}

At the core of Depminer lies its data model for dependency representation.

\subsubsection*{Location markers}
Location of each code element is represented by a \emph{location marker}, which consists of an absolute path to the file containing this element, and the range of lines in the document which a given element is encapsulating.
Some code elements, such as tokens referring to variables, will only take up a small region of code or even a fraction of a single line. 
However, some larger elements like Java methods and classes might take up a significant portion of a file, and even represent a hierarchy by containing each other.
Information on locations of code element is crucial for most applications.
For instance, knowing that a certain method lies within a certain class is important for building UML-like representations of inter-class dependencies in Java code. 

\subsubsection*{Type signatures}
To create a full representation of an element, its location data is bundled together with its \emph{type signature}.
Type signature is a description of an element derived from a field of a corresponding PSI element: it is used to classify code element from the resulted output data provided by DepMiner. 
These signatures are language specific, with concrete type labels defined by the language implementation used to build the PSI tree. 
However, in inspected cases they allow to gather important information about code elements forming a dependency. 
For instance, after a Java project is scanned, it is common to see a dependency between code elements denoted \texttt{PsiClass} and \texttt{PsiReferenceExpression}. 
In case with Java, it would convey the usage of a Java class somewhere else in the code, for instance an object of a given class is created, or a method from the class is called. 
\texttt{TypeSignature} field can be extended to support any additional language specific logic necessary in the given circumstances. 

\subsubsection*{Dependency records}
\begin{figure}[h]
    \centering
    \includegraphics[width=0.50\textwidth]{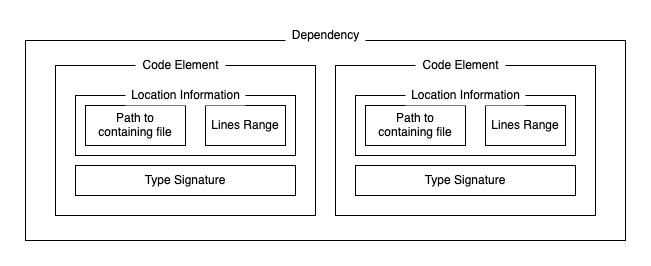}
    \caption{Representation of a dependency record in DepMiner's code.}
    \label{fig2}
\end{figure}

Location markers are combined with type signatures into a code element record. 
Similarly, two code element records are bundled together to represent a \emph{dependency record}. 
Dependencies are a final level of abstraction in DepMiner data model and the output data will be a list of dependencies.
Figure \ref{fig2} demonstrates the logical composition of a dependency record.

\subsection{Working with DepMiner}

In essence, execution of DepMiner is execution of IntelliJ IDEA, extended by DepMiner's plugin, in headless mode, i.e. without spinning up the user interface.
Headless runtime allows for pipelineable bulk processing without the need to interact with the IDE. 
In addition, headless mode improves performance by saving resources that would otherwise be occupied with loading and rendering the user interface.

DepMiner relies on Gradle for managing own dependencies and running. 
The IntelliJ Gradle Plugin~\cite{intellijgradle} is designed for building and testing IntelliJ plugins and other IntelliJ Platform based systems.
There exists a special Gradle task \texttt{runIde}, provided by the IntelliJ Gradle Plugin, that first builds and tests the plugin under development, and then launches another instance of IntelliJ IDEA with the plugin loaded. 
This allows developers to manually inspect the functionality and the UI appearance of plugin elements. 
DepMiner utilises the same \texttt{runIde} task to start an instance of IDEA and perform the analysis in headless mode.
DepMiner is launched from the command line, through the invocation of a Bash script, which starts the \texttt{runIde} task with a proper format of arguments. 

When an instance of IDEA is started, DepMiner opens and sets up a project model for the given directory, in order to trigger project indexing. 
Projects that contain files recognised by IDEs (such as .project for eclipse and .iml for IDEA) can be opened automatically by IntelliJ Platform utility functions. 
Projects that do not contain a project model definition, or store it in another format, are set up programmatically by DepMiner. \texttt{ProjectSetupUtil.kt} file in its distribution contains the implementation of programmatic project setup.

Once the indices have been built, DepMiner builds PSI trees for all files in the project. 
Traversing all of the resulting trees, each element is inspected to check if any PSI references objects are attached to the object. 
If a reference is detected, the element is added to the final analysis scope, i.e., a list of elements to be inspected further. 
The final list of PSI elements with corresponding PSI references is iterated and each reference is resolved. 
Thus, usages of code elements can be connected to their declarations. 
For each reference in the list, DepMiner constructs a \texttt{Dependency} object. 
Finally, the final list of references is produced and stored in a JSON file. 

To demonstrate how output produced by DepMiner can be potentially utilised and to support understanding of its output, we built a web app visualising
% the output produced by DepMiner 
the resulting dependency graph. 
The visualisation is available at \url{http://doi.org/10.5281/zenodo.4441366}.

\subsection{Extensibility}

DepMiner code can be easily extended in multiple ways to accommodate specific analysis needs. 
First, DepMiner can analyse source code in a variety of languages. 
To add support for additional languages, one can add a dependency on the corresponding plugin to the \texttt{build.gradle.kts}. 

Additionally, it is easy to change the scope of analysis by implementing \texttt{AnalysisScope} interface. 
Extracting dependencies is a computationally expensive task and to avoid pipeline bottlenecks one might want to exclude certain files, directories or even certain line ranges within files.

Finally, the processing logic of DepMiner can be modified to enable tasks other than building a list of dependencies. 
In fact, it can be used as a basis for tools that extract any information that is available in the scope of the IDE from arbitrary software projects.

%% file: sections/evaluation.tex
\subsection{Contribution}
\depminer is the primary contribution of this work. 
It fills the gap in the diverse field of dependency analysis tools, excluding several drawbacks of the previously mentioned, especially from a standpoint of application in research, and has a wide range of potential applications for software engineering problems.

Besides being useful as an independent data collection tool, it constitutes a a step towards reusing language analysis capabilities of IDEs for mining data for research contexts.

\subsection{Performance and Quality}
Performance of \depminer is highly influenced by the performance of IntelliJ Platform components in use. 
Constructing and analysing large PSI trees can use significant amounts of memory in large files. 
Resolving and searching for references is also computationally expensive. 
Thus, if the scope of dependency analysis can be limited, the implementation of \texttt{AnalysisScope} should strive to do so as otherwise \depminer might become a bottleneck in the analysis pipeline. 

Essential functions of \depminer are covered by functional tests. If \depminer is extended in any way (e.g. to analyse projects in other languages), the test suite should be extended accordingly. 

\subsection{Distribution}
Anonymized source code of \depminer is available on Zenodo: \url{http://doi.org/10.5281/zenodo.4441367}.